\documentclass[epj]{svjour}
\usepackage{latexsym}
\usepackage{amsmath}
\usepackage{amssymb}
\usepackage{epsfig}
\usepackage{graphicx}
\usepackage{dcolumn}
\usepackage{bm}
\usepackage{color}
\usepackage{empheq}

\definecolor{lekkoszary}{gray}{0.9}

\begin{document}

\title{Augmented Lagrangian Method for Constrained Nuclear Density Functional Theory}
\author{A. Staszczak\inst{1,2,3}  \and M.Stoitsov\inst{1,2} \and A. Baran\inst{1,2,3} \and  W. Nazarewicz\inst{1,2,4,5}}

\institute{
Department of Physics and Astronomy, University of Tennessee Knoxville, Tennessee 37996, USA
\and
Oak Ridge National Laboratory, P.O. Box 2008, Oak Ridge, Tennessee 37831, USA
\and
Institute of Physics, Maria Curie-Sk{\l}odowska University, pl. M. Curie-Sk{\l}odowskiej 1, 20-031 Lublin, Poland
\and
Institute of Theoretical Physics, Warsaw University, ul. Ho\.{z}a 69, PL-00681 Warsaw, Poland
\and
School of Engineering and Science, University of the West of Scotland, Paisley  PA1 2BE, United Kingdom
}
\date{Received: date / Revised version: date}
\abstract{The augmented Lagrangiam method (ALM), widely used in quantum chemistry constrained optimization problems, is applied in the context of the nuclear Density Functional Theory (DFT) in the self-consistent constrained Skyrme Hartree-Fock-Bogoliubov (CHFB) variant. The ALM allows precise calculations of multidimensional energy surfaces in the space of collective coordinates that are needed to, e.g., determine fission pathways and saddle points; it improves accuracy of computed  derivatives with respect to collective variables that are used to determine collective inertia; and is well adapted to supercomputer applications.
 % end of PACS codes
} % end of abstract

\maketitle

\section{Introduction}

The HFB equations of the superconducting nuclear DFT \cite{[Ben03]}
can be viewed as a constrained nonlinear optimization problem in which the total energy of the nucleus, represented by a functional of one-body densities, is minimized
subject to constraints on the  values of several independent variables. In addition to the usually imposed conditions on the number of particles (protons and neutrons), one is often interested in constraining angular momentum components (to study nuclear behavior at nonzero angular momentum) or nuclear multipole moments (or deformations) - to investigate the large amplitude collective motion, such as shape coexistence, fission or fusion.

Constrained calculations are also used when going beyond the standard single-reference DFT, e.g., within the Generator Coordinate Method \cite{[RS80]} of the multi-reference DFT \cite{[Ben08],[Dug10]}, where the constrained HFB solutions are used to generate a set of basis wave functions employed in further optimization. Another set of applications concerns the adiabatic approximation to the time-dependent HFB (ATDHFB) \cite{[RS80],[Kri74],[Bar78],[Dob81],[Rei87],[Dod00]} wherein derivatives with respect to collective coordinates  are often  approximated  by finite-difference expressions \cite{[Yul99]}.

The fission problem is of particular interest as it involves many constrained calculations along collective degrees of freedom representing families of mean fields
characterizing fission pathways and nuclear dynamics during the fission process. In particular, care should be taken to identify saddle points in a multidimensional energy surface  \cite{[Mye96],[Mam98],[Mol08a]}. In this respect, constrained calculations in many variables can be very helpful as they can separate potential energy sheets that overlap when studied in reduced-deformation spaces \cite{[Sta09]}.

An effective approach to satisfy constraints is the method of Lagrange multipliers  \cite{[Vap01]}. For example, the minimization of energy $E$ at the condition that the nuclear quadrupole moment $\hat{Q}$ has an expectation value $q^0$, is equivalent to  minimization of the Lagrangian function (or Routhian) $E'=E+\lambda (\langle\hat{Q}\rangle-q^0)$, where the Lagrange multiplier $\lambda$ is determined from the condition
\begin{equation}\label{QQ}
\langle\hat{Q}\rangle=q^0.
\end{equation}

In many cases, however,  the procedure based on a linear constraint method (LCM)  fails and the standard technique adopted is the method of quadratic  constraint  (the quadratic penalty method, QPM) \cite{Flet00,Bert99,Noce06}. In the above  example, the corresponding Lagrangian function  reads $E'=E+c (\langle\hat{Q}\rangle-q^0)^2$.
As noted in early nuclear self-consistent applications \cite{Gira70,Floc73}, results of calculations based on QPM strongly depend on the magnitude of $c$. For example, when the value of $c$ is too small, one ends up with a solution having the constrained  moment quite far away from the requested value. Increasing the value of  $c$ is often impossible as  the self-consistent procedure ceases to  converge. This is a serious deficiency of the method as  it  leaves important domains  of the collective space unresolved thus obstructing (or even preventing) the theoretical description.

An effective procedure that avoids some of the difficulties pertaining to the standard LCM but not introducing the penalty term is the  method proposed in \cite{Bass71} and used in early CHF calculations of Refs.~\cite{Cuss85,Cuss85a}, and in CHFB calculations of Refs.~\cite{[Dec80],Youn09}, in which the $\lambda$ is changed iteratively to satisfy the condition (\ref{QQ}) at each step.
Also, the constrained optimization problem is often  treated  by means of the conjugate gradient method \cite{Shew94}. However, except for a few cases \cite{[Egi95]}, most of the existing HFB solvers are based on a direct diagonalization approach and a mixing of intermediate solutions during the iteration process using a  linear  or  Broyden mixing \cite{[Bar08]}.

In this study, we demonstrate that the augmented Lagrangian method \cite{Hest69,Powe69} is an excellent alternative for nuclear-constrained HFB calculations. We  show that the method always yields self-consistent   solutions corresponding to requested values of constraints, independently of the value of the Lagrange multiplier selected. In this way, adopting the ALM, one can always access any region of  the multi dimensional energy surface  requested by the particular physical phenomena investigated. At the same time, practical  implementations of  ALM do not require more computational resources as compared to QPM.
A procedure, based on QPM but introducing a modifications of $q^{0}$ during the iterations through a  linear constraint
has been used in \cite{bonch05}. While not based on the ALM algoritm, the spirit of this method is close to ALM.

This paper is organized as follows. The method of Lagrange multipliers, in its linear and quadratic variants, is briefly discussed in Sec.~\ref{MLM} together with the  augmented Lagrangian method. The ALM algorithm adopted in this work for diagonalization-based  HFB solvers is laid out in Sec.~\ref{sec:eft5}, and the illustrative results are given in Sec.~\ref{results}. Finally, conclusions are contained in Sec.~\ref{concl}.

\section{The Method of Lagrange Multipliers}\label{MLM}
A constrained optimization problem is usually specified in terms of equality and inequality constraints \cite{Flet00,Bert99,Noce06}.
We consider here a finite-dimensional, equality-constrained non-linear optimization problem (ECP) of the form
\begin{equation}
\left\{ \begin{array}{l}
\displaystyle\min_{\boldsymbol{x}} \mathcal{E}(\boldsymbol{x})\\
\mbox{subject to } g_{i}(\boldsymbol{x})=q_{i}^{0}, \quad i=1,2,\ldots, m,
\end{array} \right.
\label{eq:1}
\end{equation}
where we assume that $\mathcal{E}: \mathbb{R}^{n}\rightarrow \mathbb{R}$ (an {\emph{objective function}}) and $g_{i}: \mathbb{R}^{n}\rightarrow \mathbb{R}$ (the {\emph{constraint functions}}) are smooth functions, and $n>m$. The Lagrangian function $E^{'}:$ $\mathbb{R}^{n+m}\rightarrow \mathbb{R}$ associated with ECP is defined as
\begin{eqnarray}
E^{'}(\boldsymbol{x},\boldsymbol{\lambda}) &=& \mathcal{E}(\boldsymbol{x})+ \sum_{i=1}^{m}\lambda_{i}[g_{i}(\boldsymbol{x})-q_{i}^{0}] \nonumber  \\
&=& \mathcal{E}(\boldsymbol{x})+ \boldsymbol{\lambda}^{T}[\boldsymbol{g}(\boldsymbol{x})-\boldsymbol{q}^{0}],
\label{eq:2}
\end{eqnarray}
where $\boldsymbol{\lambda}$=$\{\lambda_{i}\}$ is  the vector of {\emph{Lagrange multipliers}}.

The following set of necessary and sufficient conditions allow the problem  (\ref{eq:1}) to be formulated in terms of the Lagrangian function (\ref{eq:2}):
\begin{itemize}
\item The first-order necessary (\emph{local zero-slope}) condition: if $\boldsymbol{x^*}\in \mathbb{R}^n$ is a local solution of ECP (\ref{eq:1}), then there exists a unique vector $\boldsymbol{\lambda^*}\in \mathbb{R}^m$ such that $(\boldsymbol{x^*},\boldsymbol{\lambda^*})$ is a stationary point for the Lagrangian function
\begin{equation}
\boldsymbol{\nabla}_{x} E^{'}(\boldsymbol{x^{*}},\boldsymbol{\lambda^{*}})=0.
\label{eq:3}
\end{equation}
\item The second-order necessary (\emph{local convexity}) condition: if the functions $\mathcal{E}(\boldsymbol{x})$ and $g_{i}(\boldsymbol{x})$ are twice continuously differentiable, then the {\emph{Hessian}} matrix of the Lagrangian function (with respect to $\boldsymbol{x}$) must be \emph{positive semidefinite} at $(\boldsymbol{x^{*}}, \boldsymbol{\lambda^{*}})$
\begin{equation}
H(E^{'})(\boldsymbol{x^{*}}, \boldsymbol{\lambda^{*}})\succeq 0.
\label{eq:4}
\end{equation}
    In this context it should be mentioned, that Eq.~(\ref{eq:4}) is the necessary and sufficient condition for convexity of the $E^{'}(\cdot, \boldsymbol{\lambda^{*}})$ function on its convex domain.
\item The second-order sufficient conditions: assume that $\mathcal{E}(\boldsymbol{x})$ and $g_{i}(\boldsymbol{x})$ are twice continuously differentiable, and let $\boldsymbol{x^{*}}\in \mathbb{R}^{n}$ and $\boldsymbol{\lambda^{*}}\in \mathbb{R}^{m}$ satisfy the equations
\begin{equation}
\boldsymbol{\nabla}_{x} E^{'}(\boldsymbol{x^{*}},\boldsymbol{\lambda^{*}})= \boldsymbol{0},
\quad
\boldsymbol{\nabla}_{\lambda} E^{'}(\boldsymbol{x^{*}},\boldsymbol{\lambda^{*}})= \boldsymbol{0},
\label{eq:5}
\end{equation}
    and the Hessian matrix $H(E^{'})(\boldsymbol{x},\boldsymbol{\lambda})$ is \emph{positive definite} at $(\boldsymbol{x^{*}}, \boldsymbol{\lambda^{*}})$
\begin{equation}
H(E^{'})(\boldsymbol{x^{*}}, \boldsymbol{\lambda^{*}})\succ 0,
\label{eq:6}
\end{equation}
    then the vector $\boldsymbol{x^{*}}$ is a strict local solution of ECP (\ref{eq:1}).
\end{itemize}

In reality, these conditions are not always satisfied. For example, even when Eq.~(\ref{eq:1}) has a solution, the Lagrangian function $E^{'}$ could be unbounded \cite{Bass76,Bass79} since the Hessian of the Lagrangian function is not necessarily positively defined. However, it can be shown that if ECP is \emph{convex}, i.e.,   $\mathcal{E}(\boldsymbol{x})$ is a convex ($\cup$-shaped) function and $\boldsymbol{g}(\boldsymbol{x})$ is an affine function, then the local constrained minimum is unique and represents the global minimum.

Suppose that $\boldsymbol{x}(\boldsymbol{q})$ and $\boldsymbol{\lambda}(\boldsymbol{q})$ are continuously differentiable functions giving the local minima of a family of ECPs (\ref{eq:1}) parameterized by $\boldsymbol{q}\in \mathbb{R}^m$, and  $\boldsymbol{x}(\boldsymbol{q}^{0})= \boldsymbol{x^{*}}$, $\boldsymbol{\lambda}(\boldsymbol{q}^{0})= \boldsymbol{\lambda^{*}}$.
This implies \cite{Bert99} that
\begin{equation}
\boldsymbol{\nabla}_{q} \mathcal{E}[\boldsymbol{x}(\boldsymbol{q})]= -\boldsymbol{\lambda}(\boldsymbol{q})
\quad \textrm{or} \quad \boldsymbol{\nabla} E(\boldsymbol{q})= -\boldsymbol{\lambda}(\boldsymbol{q}),
\label{eq:7}
\end{equation}
where the  {\emph{primal function}} $E(\boldsymbol{q})$ is given by
\begin{equation}
E(\boldsymbol{q})\equiv \mathcal{E}[\boldsymbol{x}(\boldsymbol{q})]
= \min_{\boldsymbol{g}(\boldsymbol{x})= \boldsymbol{q}} \mathcal{E}(\boldsymbol{x}).
\label{eq:8}
\end{equation}
The interpretation  of the Lagrangian multipliers is therefore that $-\boldsymbol{\lambda}(\boldsymbol{q})$ defines the parametrical gradient of $E(\boldsymbol{q})$.
When $\mathcal{E}(\boldsymbol{x})$ is the convex function, then a primal function $E(\boldsymbol{q})$ turns out to be a convex function as well \cite{Boyd04}.

The following sections illustrate the method of Lagrangian multipliers for a one-dimensional problem. We first discuss the linear constraint method, then quadratic penalty method, and augmented Lagrangian method.

\subsection{The linear constraint method}
\begin{figure}[htb]
\centering
\includegraphics[width=0.45\textwidth]{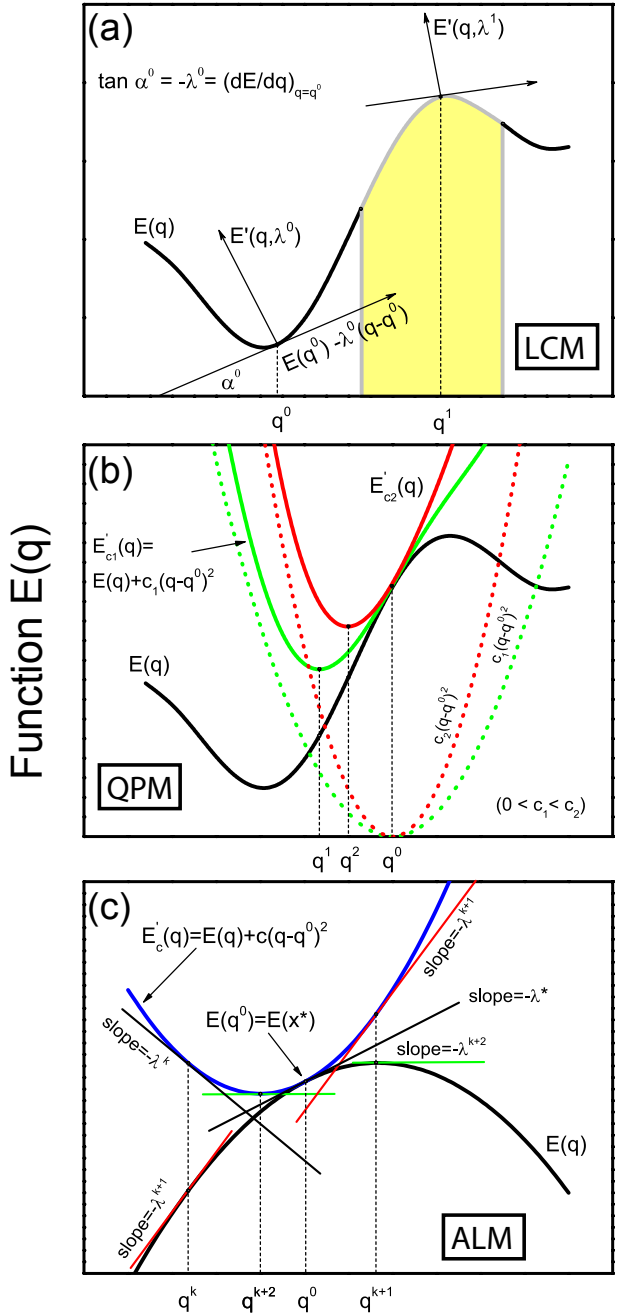}
\caption{(Color online) (a) Geometric interpretation of the linear constraint method (LCM). This method is not applicable to the shaded region, where the primal function $E(q)$ is concave. (b) Behavior of penalty function for one-dimensional QPM. (c) Geometric interpretation of the ALM iterations (based on Ref.~\cite{Bert99}). See text for  details.}
\label{fig1}
\end{figure}

The curve $E(q)$ in Fig.~\ref{fig1}(a) represents schematically how the primal function might vary as a function of $q$. It should be noted that  we analyze the primal function $E(q)$, not the objective function $\mathcal{E}(\boldsymbol{x})$.

To solve the optimization problem, one can draw the tangent to the function $E(q)$ at the point $q=q^0$, and use this line as the abscissa axis of a new coordinate system rotated by the angle $\alpha^0$ with $\tan \alpha^{0} =-\lambda^{0}= \frac{d E}{d q}(q^{0})$, see Eq.~(\ref{eq:7}). The ordinate axis in the new frame corresponds to the Lagrangian function $E^{'}(q, \lambda^{0})= E(q^{0})-\lambda^{0}(q- q^{0})$. An unconstrained minimization of $E^{'}$ gives the local minimum in the rotated system at the requested point $q^{0}$. In this way, the constrained minimization  of $E(q)$ is achieved by an unconstrained minimization of $E^{'}(q, \lambda^{0})$.

However, as discussed above, the constrained minimization procedure with LCM  can be applied only in the regions of $E(q)$ which are convex. In the concave ($\cap$-shaped) region, shaded in Fig.~\ref{fig1}(a), the function $E^{'}$ in the rotated frame has a maximum at point $q^{1}$. The minimization procedure does not yield a stable solution around the maximum, and the constrained calculation with a linear constraint function  fails to converge in the whole shaded region, see discussion in Refs.~\cite{Gira70,Floc73,Bass71} and in Sec.~7.6 of Ref.~\cite{[RS80]}.

\subsection{The quadratic constraint approach}
The local convexity assumption (\ref{eq:4}) plays a crucial part in  solving the constrained problem (\ref{eq:1}). The QPM  can be applied when the convexity of original ECP is not preserved. In such cases, one approximates the original constrained problem by an unconstrained minimization problem that involves a penalty for violation of the constraints, see Refs.~\cite{Flet00,Bert99,Noce06}, and also Refs.~\cite{[RS80],Gira70,Floc73}:
\begin{eqnarray}
\min_{\boldsymbol{x}} E^{'}_{c}(\boldsymbol{x}) &=& \min_{\boldsymbol{x}} \left\{\mathcal{E}(\boldsymbol{x})+ c \sum_{i=1}^{m}[g_{i}(\boldsymbol{x})-q_{i}^{0}]^{2}\right\} \nonumber \\
&=& \min_{\boldsymbol{x}} \left\{\mathcal{E}(\boldsymbol{x})+ c \|\boldsymbol{g}(\boldsymbol{x})-\boldsymbol{q}^{0}\|^2\right\},
\label{eq:9}
\end{eqnarray}
where $c>0$ is called \emph{penalty parameter} and $\|\cdot\|$ denotes Euclidean norm. It should be noted that if $c$ is taken sufficiently large, then the local convexity condition can be shown to hold for the Lagrangian function $E^{'}_{c}(\boldsymbol{x})$.

Figure~\ref{fig1}(b) shows the same one-dimensional case
as in Fig.~\ref{fig1}(a), but for the QPM. The primal function $E(q)$ is plotted with a solid (black) line, while the penalty function is plotted with dashed lines around the requested point $q^0$ for two different values $c_1$ and $c_2$ of the penalty parameter $c$. The resulting Lagrangian functions $E_{c}^{'}(q)=E(q)+c(q-q^{0})^{2}$
are indicated.

It is immediately seen in Fig.~\ref{fig1}(b)  that the minimum of the Lagrangian function does not correspond to  $q^0$ but rather to the values $q_1$ and $q_2$ corresponding to the penalty parameter $c_1$ and $c_2$, respectively. One can obtain the values of the function $E(q)$ in a broad range by changing the requested point $q^0$ or the penalty parameter $c$, or both. But one can neither predict in advance which value $q$ will be reached, nor to produce a regular mesh of values, which is often of interest.

We thus see that the main drawback of the QPM is that it never delivers exactly the requested constraint values. If one denotes the solution to unconstrained problem (\ref{eq:9}) by $\boldsymbol{x^{*}}(c)$ (i.e., $E^{'}_{c}(\boldsymbol{x^{*}}(c))\approx \mathcal{E}(\boldsymbol{x^{*}})=E(\boldsymbol{q}^{0})$), then it has been shown \cite{Bert99} that $\displaystyle  \lim_{c\rightarrow \infty}\boldsymbol{x^{*}}(c)= \boldsymbol{x^{*}}$. However, the Hessian matrix $H(E^{'}_{c})(\boldsymbol{x})$ is ill-defined for large values $c$. One is therefore  forced to make a compromise between satisfying the constraints and having a well-conditioned problem when using QPM \cite{Bert99,Glow95}.

\subsection{The augmented Lagrangian method}\label{SALM}

The ALM  can be viewed as a combination of LCM and QPM. It was introduced as a computational tool in 1969, by Hestenes \cite{Hest69} and Powell \cite{Powe69}, as an attempt to solve the difficulties of linear and quadratic constraint methods.

Let us begin by introducing the augmented Lagrangian function
\begin{equation}
E^{'}_{c}(\boldsymbol{x},\boldsymbol{\lambda})=
\mathcal{E}(\boldsymbol{x})+ \boldsymbol{\lambda}^{T} [\boldsymbol{g}(\boldsymbol{x})-\boldsymbol{q}^{0}]+ c \|\boldsymbol{g}(\boldsymbol{x})-\boldsymbol{q}^{0}\|^2,
\label{eq:10}
\end{equation}
which is the Lagrangian function for the problem:
\begin{equation}
\left\{ \begin{array}{l}
\displaystyle\min_{\boldsymbol{x}} \big\{\mathcal{E}(\boldsymbol{x})+ c \|\boldsymbol{g}(\boldsymbol{x})-\boldsymbol{q}^{0}\|^2\big\}\\
\mbox{subject to } \boldsymbol{g}(\boldsymbol{x})=\boldsymbol{q}^{0},
\end{array} \right.
\label{eq:11}
\end{equation}
that has the same local minima as original ECP (\ref{eq:1}). The gradient of $E^{'}_{c}(\boldsymbol{x},\boldsymbol{\lambda})$ with respect to $\boldsymbol{x}$ is
\begin{eqnarray}
\boldsymbol{\nabla}_{x} E^{'}_{c}(\boldsymbol{x},\boldsymbol{\lambda}) &=&
\boldsymbol{\nabla}\mathcal{E}(\boldsymbol{x})+ \boldsymbol{\nabla}\boldsymbol{g}(\boldsymbol{x})\big\{\boldsymbol{\lambda}+ 2c [\boldsymbol{g}(\boldsymbol{x})-\boldsymbol{q}^{0}]\big\} \nonumber\\
&=& \boldsymbol{\nabla}_{x} E^{'}(\boldsymbol{x}, \tilde{\boldsymbol{\lambda}}),
\label{eq:12}
\end{eqnarray}
where
\begin{equation}
\tilde{\boldsymbol{\lambda}}= \boldsymbol{\lambda}+ 2c [\boldsymbol{g}(\boldsymbol{x})-\boldsymbol{q}^{0}].
\label{eq:13}
\end{equation}

If $\boldsymbol{\lambda}^{k}$ is a good approximation to the solution $\boldsymbol{\lambda^{*}}$, then it is possible to approach the optimum $\boldsymbol{x^{*}}$ through the unconstrained minimization of $E^{'}_{c^k}(\,\cdot,\boldsymbol{\lambda}^{k})$ without using large values of the corresponding penalty constant $c^k$. The only condition is that  $c^k$ is  sufficiently large to ensure that the augmented Lagrangian function $E^{'}_{c^k}$ is locally convex  with respect to $\boldsymbol{x}$.

It has been shown in Refs.~\cite{Hest69,Powe69} that the use of the iterative  Lagrange multipliers
\begin{equation}
\boldsymbol{\lambda}^{k+1}= \boldsymbol{\lambda}^{k}+ 2c^{k} [\boldsymbol{g}(\boldsymbol{x}^{k})-\boldsymbol{q}^{0}]
\label{eq:14}
\end{equation}
leads to $\boldsymbol{x}^{k}$ which minimize $E^{'}_{c^k}(\,\cdot,\boldsymbol{\lambda}^{k})$.

Figure~\ref{fig1}(c) provides a geometric interpretation of the  iteration (\ref{eq:14}). To understand this figure, note that if $\boldsymbol{x}^{k}$ minimizes $E^{'}_{c^k}(\,\cdot,\boldsymbol{\lambda}^{k})$, then the vector $\boldsymbol{q}^{k}= \boldsymbol{g}(\boldsymbol{x}^{k})$ minimizes $E(\boldsymbol{q})+ (\boldsymbol{\lambda}^{k})^{T}[\boldsymbol{q}-\boldsymbol{q}^{0}]+ c^{k}\|\boldsymbol{q}-\boldsymbol{q}^{0}\|^{2}$. Hence
\begin{multline}
\boldsymbol{\nabla} \left\{E(\boldsymbol{q})+ c^{k}\|\boldsymbol{q}-\boldsymbol{q}^{0}\|^{2}\right\} \big|_{\boldsymbol{q}=\boldsymbol{q}^{k}}=\\
= \boldsymbol{\nabla}E^{'}_{c^k}(\boldsymbol{q}^{k})= -\boldsymbol{\lambda}^{k},
\label{eq:15}
\end{multline}
and
\begin{multline}
\boldsymbol{\nabla} E(\boldsymbol{q}^{k})= -\left(\boldsymbol{\lambda}^{k}+ 2c^{k}[\boldsymbol{q}^{k}- \boldsymbol{q}^{0}]\right)=\\ =-\left(\boldsymbol{\lambda}^{k}+ 2c^{k} [\boldsymbol{g}(\boldsymbol{x}^{k})- \boldsymbol{q}^{0}]\right)= -\boldsymbol{\lambda}^{k+1}.
\label{eq:15a}
\end{multline}
One can see in Fig. \ref{fig1}(c) that if $\boldsymbol{\lambda}^{k}$ is sufficiently close to $\boldsymbol{\lambda^{*}}$ and/or $c^{k}$ is sufficiently large, the next multiplier $\boldsymbol{\lambda}^{k+1}$ will be closer to $\boldsymbol{\lambda^{*}}$ than $\boldsymbol{\lambda}^{k}$ (see also Ch. 4 of \cite{Bert99}  and Ch. 17 of \cite{Noce06}).
The general iterative algorithm,  including an adjustment of $c^{k}$, can be found in  Nocedal and Wright \cite{Noce06}.

\section{The ALM Algorithm}\label{sec:eft5}

This  section provides guidance on how to implement the ALM given a working QPM algorithm (which is the standard way of carrying out  constrained minimization  with HFB solvers based on diagonalization iterations).

Let us  consider, for simplicity, a solver which defines the energy of the system $E(q)$, and $q$ is the expectation value of the (quadruple) operator $\hat{Q}$, i.e., $q=\langle\hat{Q}\rangle$.  If one wishes to compute  $E(q^0)$ for a nuclear state with a requested  quadruple deformation $q^0$,  the minimized quantity in QPM is
\begin{equation}
 E{'}(q)= E(q)+ c (q- q^{0})^2.
\label{m3}
\end{equation}
Taking the variational derivative of ({\ref{m3}}), the resulting mean-field potential can be written as
\begin{equation}
 h'= h+ 2 c (q-q^{0}) \hat{Q}.
\label{m4}
\end{equation}
In our study the penalty parameter $c$ we keep fixed. With an appropriate value of $c$, the self-consistent procedure yields a solution with quadrupole deformation, which is close to the requested value $q^0$.

Let us now apply the ALM. By taking the variational derivative of
\begin{equation}
 E{'}(q)= E(q)+\lambda (q- q^{0})+ c (q- q^{0})^2,
\label{m5}
\end{equation}
the resulting mean-field potential becomes
\begin{equation}
 h'=h+ 2 c \left(q-q^{0}(\lambda)\right) \hat{Q},
\label{m6}
\end{equation}
where $q^{0}(\lambda)=q^0-\lambda/2c$.

Comparing Eqs.~(\ref{m4}) and (\ref{m6}), we see that the use of ALM practically does not need any change in the  part of the solver that deals with constrains. One simply needs to substitute $q^0 \rightarrow q^{0}(\lambda)$. The new information that needs to be supplied is the way $\lambda$ is updated during the iteration process. According to Eq.~(\ref{eq:14}), the value $\lambda^{k+1}$  depends on the previous value $\lambda^{k}$ as:
\begin{equation}
\lambda^{k+1}= \lambda^{k}+  2 c \left(q-q^{0}\right).
\label{m7}
\end{equation}
The iterations can start from a zero value, $\lambda^{0}=0$. This is a well-defined starting point as for $\lambda=0$,  ALM reduces itself to QPM for which one assumes the solver is already working.

Comparing Eqs.~(\ref{m4}) and (\ref{m6}), one can see that in ALM  the originally requested value  $q^{0}$ is  replaced by  an effective value  $q^{0}(\lambda)$ which is adjusted during the iteration process  according to Eq.~(\ref{m7}). At the end of  iterations, of course, one ends up  with the originally requested value  $q^0$. The generalization of the ALM algorithm to the case of many constraint variables is straightforward.

\section{Results}\label{results}

The ALM algorithm of Sec.~\ref{sec:eft5} has been implemented and tested with the HFB solvers HFODD v.2.43c \cite{hfodd6} and HFBTHO \cite{hfbtho}. The illustrative examples of calculations presented below concern the spontaneous fission of $^{252}$Fm. We used the solver  HFODD, which  is capable of treating simultaneously all the possible collective degrees of freedom that might appear on the way to fission.

In the particle-hole channel, we employed the  SkM$^*$ energy density functional \cite{[Bar82]}. In the pairing channel, we adopted a seniority pairing force with the strength  parameters fitted to reproduce the experimental gaps in $^{252}$Fm \cite{[Sta07]}. The single-particle basis consisted of the lowest 1,140 stretched states originating from the lowest 31 major oscillator shells. The details of the calculations can be found in Ref.~\cite{[Sta09]}. We wish to remark only  that special care should be taken when combining ALM with the Broyden mixing \cite{[Bar08]}.

\begin{figure}[htb]
\centering
\includegraphics[width=0.45\textwidth]{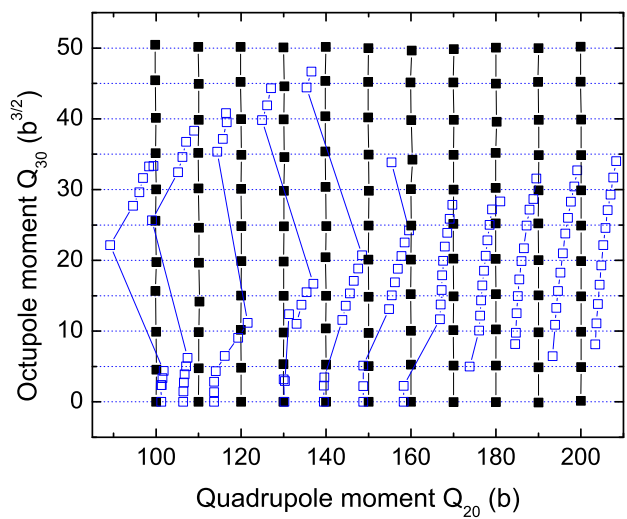}
\caption{(Color online)
A comparison between the ALM (black squares) and the standard QPM (open squares)  for the constrained self-consistent convergence scheme. The HFB calculations were carried out for the total energy surface  of $^{252}$Fm in a two-dimensional plane of elongation, $Q_{20}$, and reflection-asymmetry, $Q_{30}$. Although QPM often fails to produce a solution at the required values of constrained variables ($Q_{20}$, $Q_{30}$) on a rectangular grid, ALM performs very well in all cases.
}
\label{fig4}
\end{figure}

Figure~\ref{fig4} shows the results of constrained DFT calculations on a two-dimensional Cartesian grid of quadrupole ($Q_{20}$) and octupole ($Q_{30}$) moments. The requested values of constrained moments correspond to the grid points. While the ALM yields these values very precisely, the standard QPM
yields multipole moments that are often very different from the requested ones. In particular, QPM is unable to cover the
whole area of interest.

\begin{figure}[htb]
\centering
\includegraphics[width=0.48\textwidth]{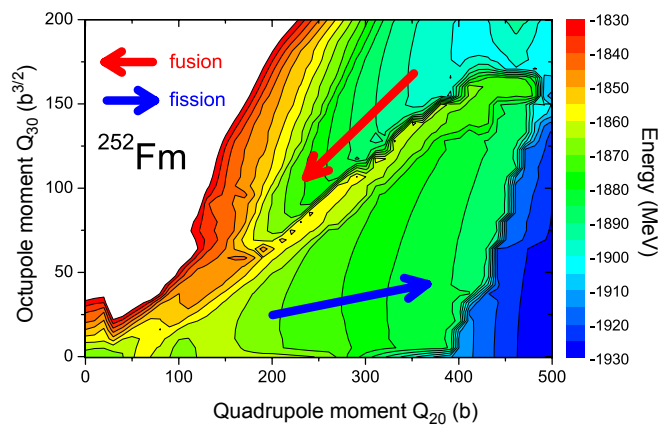}
\caption{(Color online) Two-dimensional total energy surface for $^{252}$Fm calculated with SkM$^*$ energy density functional
using the ALM in the
plane of  collective coordinates $Q_{20}$-$Q_{30}$. The fission
and fusion pathways are marked.
The difference between contour lines is 5\,MeV.}
\label{fig5}
\end{figure}

Figure~\ref{fig5} shows the  total energy surface of $^{252}$Fm in the $Q_{20}$-$Q_{30}$ plane obtained with ALM. Comparing with Fig.~\ref{fig4}, one can see interesting physics in the region which was inaccessible by QPM, namely the appearance of the second (fusion) valley at large values of $Q_{30}$ separated from the spontaneous fission valley by a steep ridge.

\section{Conclusions}\label{concl}

The augmented Lagrangiam method to solve a constrained nonlinear  problem of CHFB has been  compared with the standard variants of the method of Lagrange multipliers, i.e., quadratic penalty method and linear constraint method. We discuss the numerical strategy beyond QPM and ALM algorithms and show how to implement ALM in HFB solvers based on the diagonalization approach.

Compared to QPM, we find  ALM to be  superior: it enables precise constrained calculations  in many dimensions thus uncovering regions of collective space that are not accessible with the standard method. The method  is well adapted to supercomputer applications and its
stability  makes it a tool of choice for  large-scale  CHFB calculations, such as computations of multidimensional fission pathways discussed in this work.

\section*{Acknowledgments}
We wish to thank Robert Harrison and Arthur Kerman for useful suggestions and comments.
This work was supported by the U.S. Department of Energy under
Contract Nos. DE-FC02-09ER41583 (UNEDF SciDAC Collaboration),
DE-FG02-96ER40963 (University of Tennessee);
by the National Nuclear Security Administration under the Stewardship Science Academic Alliances program through DOE Grant
DE-FG52-09NA29461; by the NEUP grant
DE-AC07-05ID14517 (sub award
00091100);
by the Polish Ministry of Science and Higher Education - Contract
N~N202~231137.
Computational resources were provided by the National Center
for Computational Sciences at Oak Ridge National Laboratory.

%\bibliographystyle{unsrt}
%\bibliography{jacwit26,bibalm}

\end{document}